# Accessing negative Poisson's ratio of graphene by machine learning interatomic potentials

Jing Wu[1], E Zhou[1], Zhenzhen Qin[2], Xiaoliang Zhang[3,*], and Guangzhao Qin[1,*]

[1]*State Key Laboratory of Advanced Design and Manufacturing for Vehicle Body, College of Mechanical and Vehicle Engineering, Hunan University, Changsha 410082, P. R. China*

[2] *International Laboratory for Quantum Functional Materials of Henan, and School of Physics and Microelectronics, Zhengzhou University, Zhengzhou 450001, China*

[3] *Key Laboratory of Ocean Energy Utilization and Energy Conservation of Ministry of Education, School of Energy and Power Engineering, Dalian University of Technology, Dalian 116024, China*

**Abstract**:

The negative Poisson's ratio (NPR) is a novel property of materials, which enhances the mechanical feature and creates a wide range of application prospects in lots of fields, such as aerospace, electronics, medicine, *etc*. Fundamental understanding on the mechanism underlying NPR plays an important role in designing advanced mechanical functional materials. However, with different methods used, the origin of NPR is found different and conflicting with each other, for instance, in the representative graphene. In this study, based on machine learning technique, we constructed a moment tensor potential (MTP) for molecular dynamics (MD) simulations of graphene. By analyzing the evolution of key geometries, the increase of bond angle is found to be responsible for the NPR of graphene instead of bond length. The results on the origin of NPR are well consistent with the start-of-art first-principles, which amend the results from MD simulations using classic empirical potentials. Our study facilitates the understanding on the origin of NPR of graphene and paves the way to improve the accuracy of MD simulations being comparable to first-principle calculations. Our study would also promote the applications of machine learning interatomic potentials in multiscale simulations of functional materials.

*Author to whom all correspondence should be addressed.
E-Mail: X.Z. <zhangxiaoliang@dlut.edu.cn>, G.Q <gzqin@hnu.edu.cn>

# Introduction

The Poisson's ratio is a basic property that characterizes the deform behavior of materials under tension or stress. Normally, a contractive phenomenon appears in the transverse direction when it is stretched along the longitude direction, which is named as the positive Poisson's ratio. However, a minority of materials with novel mechanical properties could show expansive behavior, which is called as negative Poisson's ratio (NPR). The distinct trait has enormous prospects for optimizing the mechanical property, which bring a research focus[1–8]. Among the systems with NPR, the study of NPR phenomena in graphene is especially prominent[9–16]. Some representative approaches have been employed to modulate the NPR behavior, such as uniaxial tension[15], defects[17] and hydrogenation[10]. In literature, the NPR of monolayer graphene has been studied using molecular dynamics (MD) simulation[9,14,18] and first-principle calculation based on density functional theory[19] (DFT). For instance, using MD simulation, Jiang *et al.*[18] reported the NPR in graphene, and revealed that the NPR is resulted from the increase of bond length. On the contrary, by DFT calculations, Qin *et al*[19] found the increase of bond angle instead of bond length is responsible for NPR in graphene. It was analyzed that the discrepancy lies in the inaccuracy of the classic empirical potential used in the MD simulations. Thus, it is of urgent demand to develop a potential for graphene of high accuracy, which is of great significance to the multiscale simulations of graphene-based mechanical functional applications.

Recently, with the rapid development of artificial intelligence (AI) technology based on machine learning, more and more researchers use AI technology to construct interatomic potentials for two-dimensional (2D) materials and further complex compounds[20–30], which display high accuracy compared to the classical empirical potentials. Currently, lots of AI technology based methods have been employed to generate interatomic potential, such as Gaussian approximation potentials[31] (GAP), Spectral neighbor analysis potential[32,33] (SNAP), Artificial neural networks[34] (ANN), and Moment tensor potentials[35,36] (MTP). It has been comparatively shown in literature that the MTP possesses lots of advantages, such as extremely high accuracy, high transferability, and low computation costs[28,37–40].

In this work, by adopting the Moment Tensor method, we generated a machine learning interatomic potential for accurately describing the mechanical properties of graphene, with the aim of accessing the origin of NPR. To validate the accuracy of the developed MTP, the phonon spectrum, heat capacity and the change of potential energy are calculated in comparison with the results of DFT and few commonly used classic empirical potentials. With uniaxial strain applied, we systematically analyzed the evolution of the key geometry parameters for understanding why NPR emerges in graphene. It is found that the increase of bond angle causes the NPR of graphene, which is well consistent with DFT calculations. The developed MTP in this study would shed light on the applications of machine learning interatomic potentials in multiscale simulations of functional materials.

## The development of the MTP

All the first-principles calculations were performed using the Vienna Ab-initio Simulation Package[41] (VASP). The Perdew–Burke–Ernzerhof[42] functional of generalized gradient approximation was introduced in the calculations. For the *ab-initio* calculations, the kinetic energy cutoff was set as 1000 eV, a 15 × 15 × 1 Monkhorst-Pack[43] *k*-point grid is employed. To train a machine learning interatomic potential model, the required data sets were built by *ab-initio* MD (AIMD) simulations. These necessary AIMD simulations were performed with a 3 × 3 × 1 *k*-mesh under 600 eV kinetic energy cutoff. To diversify the data sets, the simulation was carried out at different temperatures of 100, 300, 500, and 700 K, respectively. The 2000 time steps simulations are conducted for each temperature.

Compared to classical empirical potentials, the MTP parameters were trained passively using the configurations provided by AIMD simulations. The training process can be treated as performing an energy minimization of the system, as depicted in the following formula.

$$\sum_{k=1}^{k} \left[ w_e \left( E_k^{\text{DFT}} - E_k^{\text{MTP}} \right)^2 + w_f \sum_i^N \left| f_{k,i}^{\text{DFT}} - f_{k,i}^{\text{MTP}} \right|^2 + w_s \sum_{i,j=1}^{3} \left| \sigma_{k,ij}^{\text{DFT}} - \sigma_{k,ij}^{\text{MTP}} \right|^2 \right] \to \min, \quad (1)$$

, where $E_k^{\text{DFT}}$, $f_{k,i}^{\text{DFT}}$ and $\sigma_{k,ij}^{\text{DFT}}$ are the energy, force and stress calculated by DFT, respectively. $E_k^{\text{MTP}}$, $f_{k,i}^{\text{MTP}}$ and $\sigma_{k,ij}^{\text{MTP}}$ are the energy, force and stress predicted by MTP, respectively, where *k* and *N* are the number of configurations from AIMD and the number of atoms in the studied system, respectively. The $w_e, w_f$ and $w_s$ are the corresponding weight value, which were set to 1, 0.1, 0.001 in this work. First, considering the time and computation costs, the initial training sets only include one-tenth of overall data sets. Then, we update the training sets with a reliable extrapolations grade[39]. With the updated data, the MTP was retrained.

Based on the optimized MTP, molecular dynamics simulation was performed using the Large-scale Atomic/Molecular Massively Parallel Simulator (LAMMPS) software package[44]. For comparison, the Carbon-based AIREBO[45], REBO[46] and Tersoff[47] classical empirical potentials were also introduced to perform the MD simulations using LAMMPS.

## The validation of the developed MTP

To analyze the accuracy and transferability of our machine learning model of the developed MTP, we perform an examination on the deviation of the energy and force predicted by MTP from first principles. Fig. 1 shows the comparison of energies and forces along three directions predicted by the developed MTP and the DFT calculations. It is worthwhile to note that all the results predicted by the MTP are in good agreement with DFT calculations. Notably, the three directions force value predicted by machine learning interatomic potential is well consistent with the DFT calculations as shown in Fig. 1(b), (c) and (d). The root mean square error (RMSE) of forces predicted by MTP are 0.0124, 0.0126, and 0.0008 eV/Å on X, Y, and Z directions, respectively. Thus, the performance of the MTP developed in this study are superior to the results reported by Patrick Rowe *et al.*[22], where the RMSE in force is reported to be 0.028 eV/Å using the GAP method.

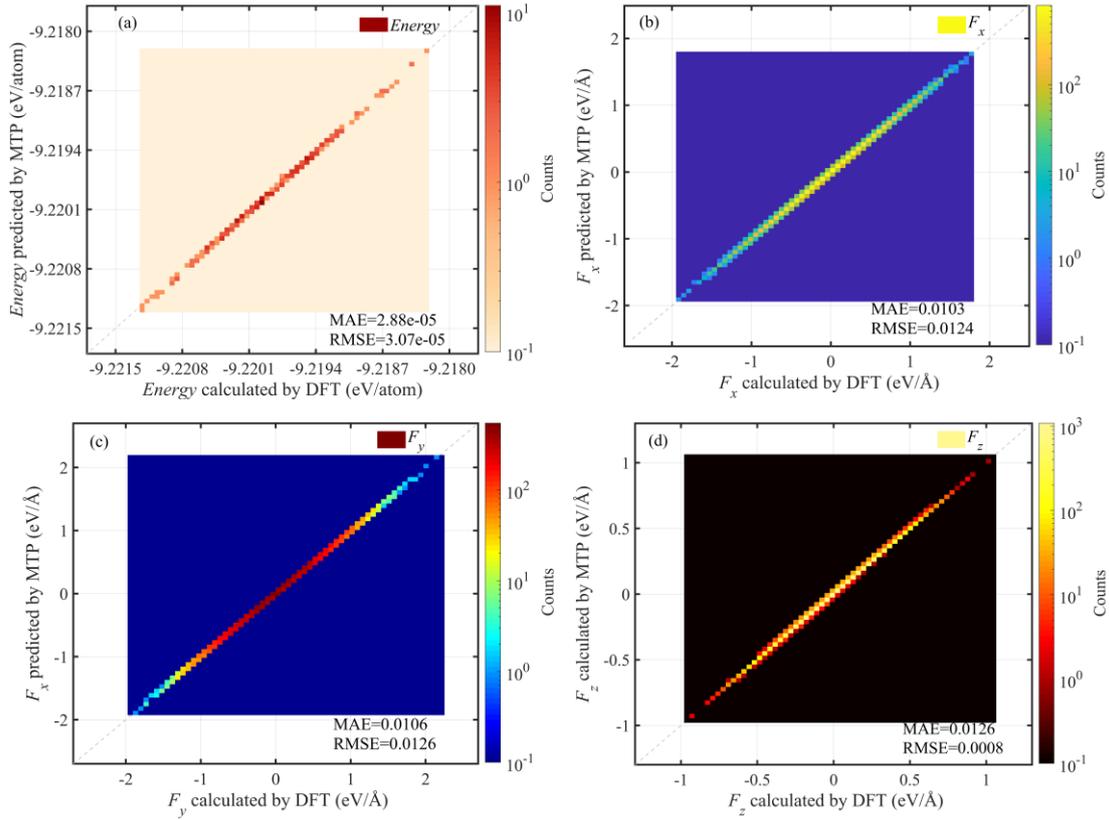

Fig. 1 The comparison of (a) atomic energies and forces (b, c, d) along three directions (*x, y, z*) between MTP predictions and DFT calculations. The color indicates the number density of points.

It is well known that the ability of describing lattice dynamics is extremely significant for evaluating the performance of interatomic potential model. Thus, to further test the accuracy and reliability of the developed MTP, we calculated the phonon spectrum of graphene describing atomic vibrations. As shown in Fig. 2(a), the phonon spectrum calculated by the MTP is directly compared to the results calculated by DFT and the representative classical interatomic potential of AIREBO. The overall performance of the developed MTP is pretty good in excellent agreement with the DFT calculations. Specifically, the low and high frequency branches are also precisely predicted by the MTP. It is worth noting that the performance of classical interatomic potential is not satisfied, especially for the high frequency optical phonon branches. The obviously overestimated high frequency phonon branches by the AIREBO are also revealed by the density of states (DOS).

Additionally, Fig. 2(b) shows the comparative results of energy strain curves, indicating the developed MTP in this study can capture the mechanical trait of graphene. Using the finite displacement method, the heat capacity $C_V$ is also calculated as shown in Fig. 2(c). Both the MTP and AIREBO are good at describing $C_V$, while the result obtained by MTP is closer to the DFT results. In short, our machine learning interatomic potential of the MTP is capable to accurately describing a variety of attributes of graphene, which shows significant advantages compared to the classical empirical potentials.

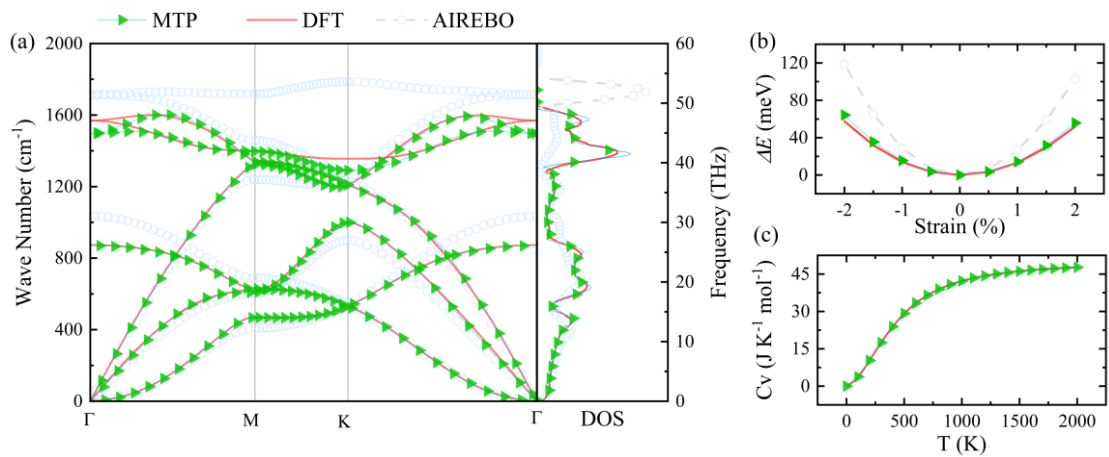

Fig. 2 The comparison among MTP, DFT and AIREBO. (a) The phonon spectrum and the density of states. (b) The change of potential energy against strain. (c) The heat capacity as a function of temperature.

## The NPR in graphene

The NPR is a mechanical phenomenon, which mainly depends on the interatomic interactions. It has been verified in Figs. 1 and 2 that our machine learning interaction potential is capable to accurately capture the atomic force feature in graphene. In the following, the NPR phenomenon in graphene are investigated with uniaxial strain applied along armchair and zigzag directions, respectively. The corresponding response of the driven strain are presented in Fig. 3. Commonly, when it is stretched by tensile strain along one direction, the lattice constant of another orthogonal direction will decrease. However, for graphene, when uniaxial strain is applied to the armchair direction and exceeds a threshold, the zigzag direction shows expansive behavior instead of contraction. As shown in Fig. 3(a), there is an obvious valley, indicating the Poisson's ratio transiting from positive to negative. In contrast, when strain is applied to zigzag direction, the special phenomenon does not arise [Fig. 3(b)]. Note that for the classical empirical potential of Tersoff, an anomalous behavior emerges, which is different from the results of MTP, DFT and other classical empirical potential. Such an anomalous behavior has been also reported by Fan et al[14], which may lie in the inaccurate description of the interatomic interactions by the Tersoff potential.

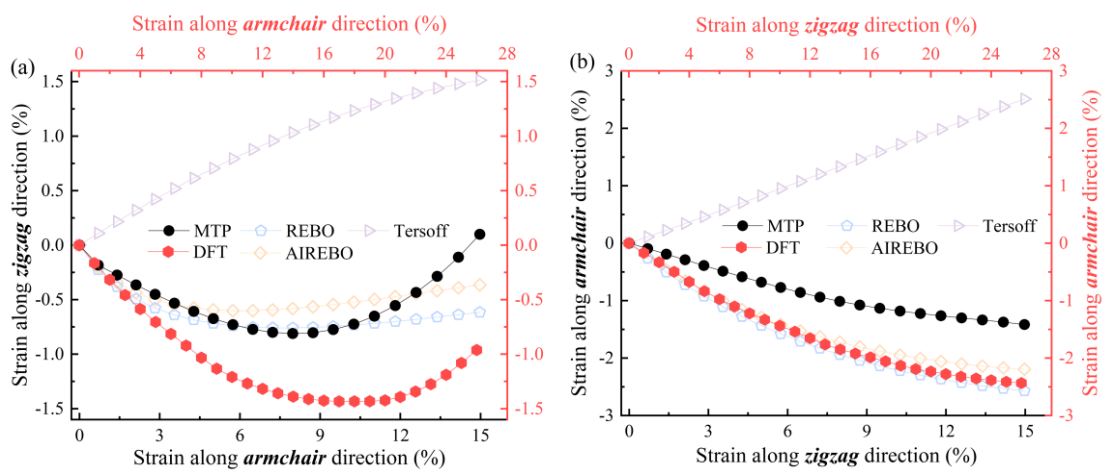

Fig. 3 The response of the driven strain under uniaxial strain along (a) armchair and (b) zigzag directions, respectively. The results are calculated by MTP, DFT and classical empirical potentials of REBO, AIREBO, and Tersoff. The red line (with red axis) denotes the results calculated by DFT, and the other lines (with black axis) denote the results calculated by MD (the following figures are also identical).

The Poisson's ratio is defined as $\upsilon = -\varepsilon_y/\varepsilon_x$, where $\varepsilon_x$ and $\varepsilon_y$ are the applied strain and the resultant strain, respectively. The strain can be gained by $\varepsilon = (l-l_0)/l_0$, where $l$ and $l_0$ are the deformed and initial lengths, respectively. In the following, we mainly focus on the NPR situation with strain applied along the armchair direction. The applied uniaxial strain along armchair direction is up to 15% and 28% (before the failure of graphene structure) for MD simulations and DFT calculations, respectively. The Fig. 4 displays the evolution of Poisson's ratio with the uniaxial strain along armchair direction. When the strain over a certain threshold value the NPR behavior emerges. It is observed that the NPR occurred around 6% strain for the REBO and AIREBO empirical potentials, which is in good agreement with previous reports[18]. The Tersoff empirical potential describes an intrinsic NPR behavior at overall strain range. As for the machine learning interatomic potential, the predicted threshold value for NPR emerging is around 9%, which is larger than the classical empirical potentials and smaller than the DFT calculations. The predicted NPR value (-0.261) is larger compared to DFT calculation (-0.156) and MD simulation (-0.028) at ultimate strain. Despite the slight difference, the overall evolution tendency of NPR phenomenon predicted by the machine learning interatomic potential is identical to the DFT calculation.

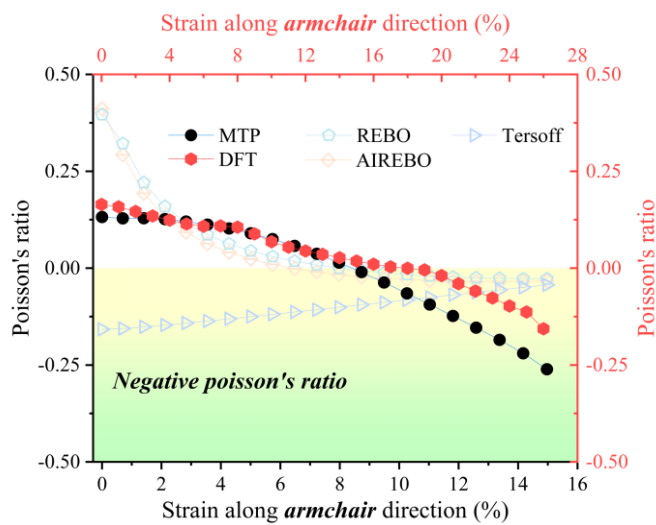

Fig. 4 The Poisson's ratio calculated by MTP, DFT and few other classic empirical potentials (REBO, AIREBO, Tersoff) under uniaxial strain along armchair direction.

Although lots of literatures have reported the NPR phenomenon[18,19], there is an unsolved discrepancy on the fundamental origin between classical empirical potentials and DFT calculations. For instance, the NPR has been widely accepted to be caused by the competition among bond angle ($\theta$) and bond length ($b_1$). Jiang et al[18] reported that the increase of bond length is responsible for the NPR based on the MD simulation with empirical Brenner potential. However, based on DFT calculation, Qin et al[19] revealed that the underlying mechanism lies in the variation of bond angle instead of bond length, which is in sharp contrast to the MD simulations with classical empirical potentials.

To fundamentally understand the underlying mechanism of NPR in graphene, we performed a comparative study on the evolution of key geometric parameters as depicted in Fig. 5. The auxetic property of graphene arises from the increasing lattice constant along zigzag direction, which is governed by the variation of bond angle $\theta$ and bond length $b_1$ [Inset of Fig. 5(a)]. With tensile strain applied along armchair direction, the resultant tendencies of MTP and DFT are nearly identical, which are opposite to the results obtained from classical empirical potentials. Fig. 5(b) shows that the angle $\theta$ predicted by MTP and DFT calculations first decreases and then increases, which is distinct from classical empirical potentials that bond angle keeps decreasing. For bond length $b_1$, the result of DFT and MTP also present a similar phenomenon that the bond length first increase and then decrease. In contrast, the bond length calculated by classical empirical potentials continually increases with the increasing tensile strain. The opposite variation of bond length and bond angle between (DFT, MTP) and the classical empirical potentials indicates the different mechanism underlying the NPR. For DFT and MTP calculation, when $\theta$ starts to increase, $b_1$ decreases due to the decreased projection force. Consequently, the increase of $\theta$ is responsible for NPR. However, the bond angle continually decreases for classical empirical potential calculation, and the stretching force projected on the bond increases. Thus, the increasing bond length results in the NPR. The difference in the underlying mechanism is caused by the difference in the description of the interatomic interactions. The DFT calculations naturally involves the effect of the evolution of electronic structures with strain applied, and thus is more accurate than the classical empirical potentials. Although the developed MTP in this work is trained based on the data generated in the cases with no strain, it successfully captures the effect of the evolution of electronic structures with strain applied and shows the power to effectively reveal the mechanism underlying the NPR phenomenon.

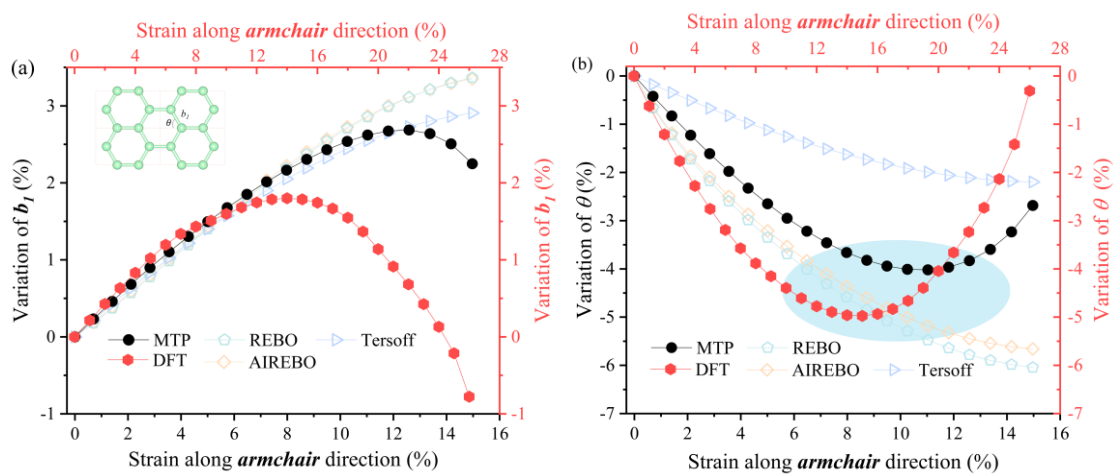

Fig. 5 The evolution of key geometry parameters of (a) $b_1$ and (b) $\theta$, which are labeled on the graphene geometry structure as shown in the inset of (a). The colored ellipse in (b) marks the significant increase of the $\theta$.

# Conclusion

In summary, we have successfully applied the MTP model to generate a machine learning interatomic potential for graphene. The developed machine learning potential well captures the energies and forces of graphene with low RMSE compared to the state-of-art DFT calculations. To further benchmark the quality of the developed MTP, we performed a systematical study on the NPR phenomena of graphene with comparison to few commonly used classic empirical potentials and the DFT calculations. The results show that the MTP can precisely capture the features of phonon spectrum, the change of potential energy against strain, and the heat capacity against temperature. With the trained MTP, the NPR phenomenon of graphene is well reproduced. By analyzing the evolution of key geometries, the increase of bond angle is found to be responsible for the NPR instead of the variation of bond length. The results amend the discrepancy from MD simulation using classic empirical potentials. Our study paves the way to improve the accuracy of MD simulations by extending the applications of machine learning interatomic potentials. For instance, in literature, a machining learning interatomic potential is mainly trained based on complex training sets. However, the MTP developed in this study is trained with simple and no strain cases, which means the MTP possesses high transferability. With the higher computational efficiency compared to the DFT calculations, the machine learning interatomic potential of MTP provides new opportunity for the multiscale simulations of functional materials.

# Data availability

The datasets (MTP potential, training set, and LAMMPS input file) in this study are available at https://github.com/WU2JING/NPR .

# Acknowledgement

We acknowledge the support of the National Natural Science Foundation of China (Grant Nos. 52006057, 52076031, 11904324), the Fundamental Research Funds for the Central Universities (Grant Nos. 531118010471 and 541109010001), and the Changsha Municipal Natural Science Foundation (Grant No. kq2014034). The numerical calculations in this paper have been done on the supercomputing system of the National Supercomputing Center in Changsha.